\documentclass[prb,twocolumn,amsmath,amssymb,floatfix,superscriptaddress]{revtex4-1}
\usepackage{graphicx}
\usepackage[FIGTOPCAP]{subfigure}
\usepackage[usenames]{color}
\usepackage{hyperref}

\begin{document}
\title{Manipulating superconductivity in ruthenates through Fermi surface engineering}
\date{\today}
\author{Yi-Ting Hsu}
\affiliation{Department of Physics, Cornell University, Ithaca, New York 14853, USA}
\author{Weejee Cho}
\affiliation{Department of Physics, Stanford University, Stanford, California 94305-4060, USA}
\author{Bulat Burganov}
\affiliation{Department of Physics, Cornell University, Ithaca, New York 14853, USA}
\author{Carolina Adamo}
\affiliation{Department of Materials Science and Engineering, Cornell University, Ithaca, New York 14853, USA}
\affiliation{Department of Applied Physics, Stanford University, Stanford, CA 94305, USA}
\author{Alejandro Federico Rebola}
\affiliation{School of Applied and Engineering Physics, Cornell University, Ithaca, New York 14853, USA}
\author{Kyle M. Shen}
\affiliation{Department of Physics, Cornell University, Ithaca, New York 14853, USA}
\affiliation{Kavli Institute at Cornell for Nanoscale Science, Ithaca, New York 14853, USA}
\author{Darrell G. Schlom}
\affiliation{Department of Materials Science and Engineering, Cornell University, Ithaca, New York 14853, USA}
\affiliation{Kavli Institute at Cornell for Nanoscale Science, Ithaca, New York 14853, USA}
\author{Craig J. Fennie}
\affiliation{School of Applied and Engineering Physics, Cornell University, Ithaca, New York 14853, USA}
\author{Eun-Ah Kim}
\affiliation{Department of Physics, Cornell University, Ithaca, New York 14853, USA}

\begin{abstract}
The key challenge in superconductivity research is to go beyond the historical mode of discovery-driven research. 
We put forth a new strategy, which is to combine theoretical developments in the weak-coupling renormalization group approach with the experimental developments in lattice strain driven Fermi surface-engineering. 
 For concreteness we theoretically investigate how superconducting tendencies will be affected by strain engineering of ruthenates' Fermi surface. We first demonstrate that our approach qualitatively reproduces recent experiments under uniaxial strain. We then note that order few $\%$ strain readily accessible to epitaxial thin films,  can bring the Fermi surface close to van Hove singularity. Using the experimental observation of the change in the Fermi surface under biaxial epitaxial strain and ab-initio calculations, we  predict ${\rm T}_{\rm c}$ for triplet pairing to be maximized by getting close to the 
van Hove singularities without tuning on to the singularity. 
\end{abstract}

\maketitle

\textit{Introduction} $-$ The notion of topological superconductivity \cite{raeyedge,PhysRevB.61.10267,PhysRevLett.86.268,PhysRevB.73.220502,RevModPhys.80.1083,PhysRevB.69.184511,PhysRevLett.99.197002}  drove intense investigation of a triplet superconductor
$\rm{Sr_2RuO_4}$ \cite{Ishida1998,Nelson12112004,Kidwingira24112006,PhysRevB.76.014526,PhysRevB.89.144504,Jang14012011}.
Unfortunately its fairly low transition temperature $\rm{T_c}\sim$ 1.5 K\cite{SROscDisorder} has been one of the limiting factors for experimental studies. 
Naturally there has been much interest in enhancing the $\rm{T_c}$ of $\rm{Sr_2RuO_4}$. Since the $\rm{T_c}$ is extremely sensitive to disorder, the usual tuning knob of doping is not an option. On the other hand, successes in both local enhancement of $\rm{T_c}$ in eutectic samples\cite{PhysRevLett.81.3765} and near dislocations\cite{Ying2013} and in global enhancement of $\rm{T_c}$ using c-axis uniaxial pressure\cite{doi:10.1143/JPSJ.78.103705,PhysRevB.81.180510} and in-plane uniaxial strain\cite{Hicks18042014} point to a new knob: the lattice strain. Now the key question is how to connect this new knob to a theoretical framework that can guide the quest for higher $\rm{T_c}$ topological superconductor. 

$\rm{T_c}$ is generally hard to theoretically predict since it is a non-universal quantity which depends on microscopic details of the system. 
The fact that one cannot just apply mean-field theory for repulsion-driven anisotropic superconductors makes it even worse. 
Nevertheless \textcite{KL} have observed early on
that even with a short-range bare repulsion, the momentum-dependence in the irreducible particle-particle vertex from higher order corrections can still give rise to a Cooper instability in a suitable channel. This insight was further developed for Hubbard type models on lattice \cite{RGSchulz2,fRGMetzner,SalmhoferRG,fRGRice,PhysRevB.81.224505}. A common thread in these approaches is the fact that the band structure near Fermi surfaces (FS) determines 
the bare susceptibilities which enter the expression for the pairing interaction. 
This invites the notion of controlling superconductivity through controlling fermiology, going beyond the traditional approach of doping \cite{AFscPnictidesLee,DopedpnictidesThomale,IronScChubukov,Nandkishore2012}.

Our idea is to employ the weak-coupling renormalization group (RG) approach\cite{PhysRevB.81.224505,PhysRevLett.105.136401} in embracing the new experimental knob of lattice strain. 
Since the pioneering work of Chu et al\cite{Chu710}, piezoelectric-based control of lattice strain has become a new knob. This approach was further developed\cite{piezoHicks} to enable substantial uniaxial strain on bulk Sr$_2$RuO$_4$ and led to a 40$\%$ enhancement of $T_c$\cite{Hicks18042014}. 
A recent experimental advance by some of us in growing epitaxially strained ruthenate films\cite{Burganov2015} presents a new opportunity. 
This is particularly exciting because the epitaxial strain can
dramatically alter the band structure\cite{Burganov2015}. 

Here we theoretically investigate how strain affects the fermiology and the associated superconducting tendencies.  For this we extract tight-binding parametrization from angle-resolved photoemission spectroscopy (ARPES) data and 
a density functional theory(DFT) calculation on strained systems. We then use the tight-binding model as the microscopic input to the RG calculation to study superconducting instability. We examine the cases of piezoelectric-based uniaxial strain\cite{Hicks18042014} and epitaxial biaxial strain\cite{Burganov2015}. We reproduce the observed trend for the case of uniaxial strain and predict 
non-monotonic dependence of the $\rm{T_c}$ on the biaxial strain.

\textit{The Model and the Approach} $-$
Our microscopic starting point is a three-band Hubbard model derived from the Ru $t_{2g}$ orbitals $d_{xz}$, $d_{yz}$, and $d_{xy}$:
\begin{align}
H= \sum_{\vec{k}\alpha\sigma}\epsilon^{\alpha}(\vec{k})c^{\dagger}_{\vec{k},\alpha,\sigma}c_{\vec{k},\alpha,\sigma}
+U\sum_{i\alpha}n_{i,\alpha,\uparrow}n_{i,\alpha,\downarrow},
\label{eq:bareH}
\end{align} 
where $\vec{k}=(k_x,k_y)$, $\alpha=xz,yz,xy$, $\sigma=\uparrow,\downarrow$ denote the crystal momentum, the orbital index, and the spin respectively, and $n_{i,\alpha,\sigma}\equiv c^{\dagger}_{i,\alpha,\sigma}c_{i,\alpha,\sigma}$.
Given the well-established unconventional aspects of superconductivity in bulk  $\rm{Sr_2RuO_4}$ \cite{Ishida1998,Nelson12112004,Kidwingira24112006}, we focus on the 
repulsive intra-orbital on-site repulsion $U>0$\cite{KL,PhysRevLett.105.136401}. 
\footnote{We are not including the inter-orbital repulsion $V$ in this letter. Nevertheless we have checked that the inter-orbital $V\leq 0.5U$ makes no qualitative difference to the results we report.}

For the intra-orbital kinetic energies $\epsilon^{\alpha}(\vec{k})$ we employ the following tight-binding parameterization:
\begin{align}
&\epsilon^{xz}(\vec{k})=-2t_x\cos k_{x}-2t^{\perp}_y\cos k_{y}-\mu_1\nonumber\label{eq:epsilon}\\
&\epsilon^{yz}(\vec{k})=-2t_y\cos k_{y}-2t^{\perp}_x\cos k_{x}-\mu_1\nonumber\\
&\epsilon^{xy}(\vec{k})=-2(t'_x\cos k_x+t'_y\cos k_y)-4t''\cos k_x\cos k_y-\mu_2
\end{align}
where we neglect the orbital-mixing terms. Although
\textcite{PhysRevB.89.220510} found the spin-orbit coupling in particular to significantly alter the nature and mechanism of pairing in the unstrained system, the van Hove singularities occur at point $X=(\pi,0)$ and $Y=(0,\pi)$ which lie in the region of the FS where orbital characters are well defined\cite{SOCMackenzie,SROreviewKallin}
Hence we expect the absence of orbital-mixing terms in our model would not affect our conclusions in a qualitative manner. 
The dispersions of the three bands in Eq.~\eqref{eq:epsilon} yield two quasi-one dimensional(1D) FS's  consisting of the Ru orbitals $d_{xz}$ and $d_{yz}$, and
one quasi-two dimensional(2D) FS consisting of the 
Ru orbital $d_{xy}$. 
\begin{figure}[t]
	\includegraphics[width=.4\textwidth]{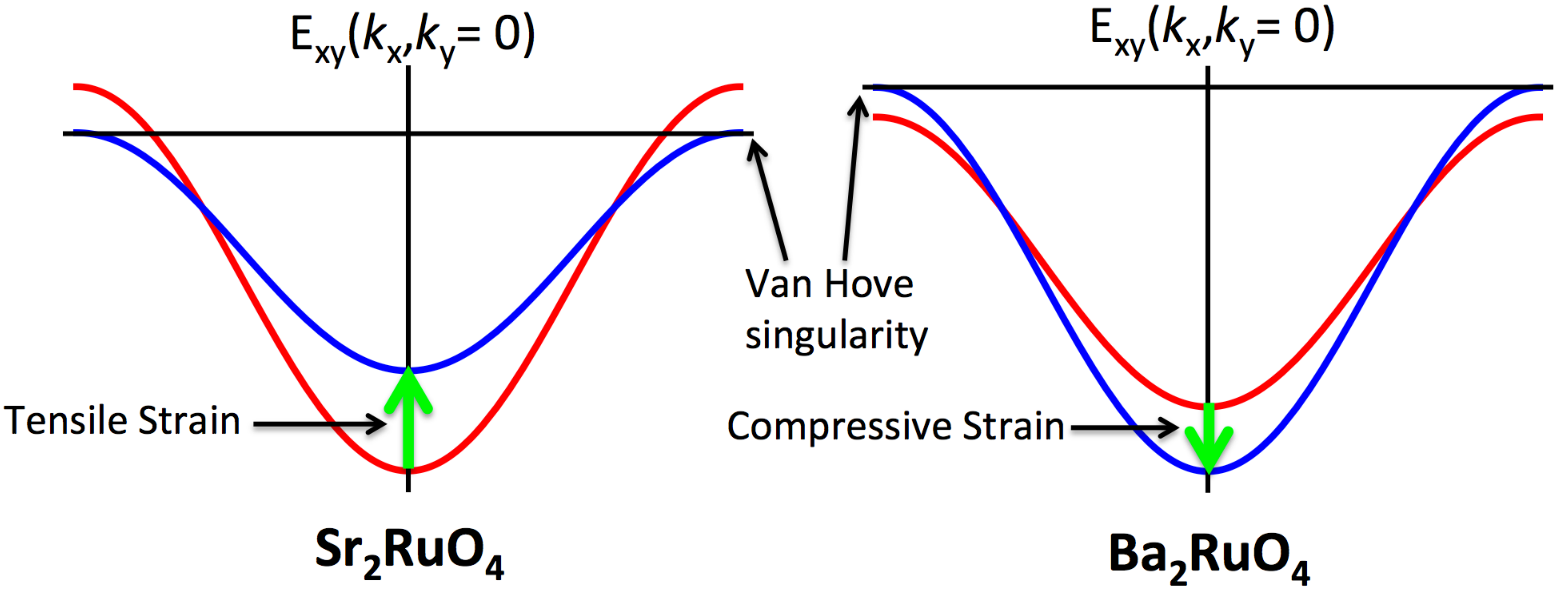}
\caption{The effect of epitaxial biaxial strain on the $xy$ 2D-band in $\rm{Sr_2RuO_4}$ and $\rm{Ba_2RuO_4}$, respectively. Red curves are unstrained bands. Bands were obtained by fitting tight-binding parameters to DFT data.}
\label{bands_strain}
\end{figure}

We connect the lattice strain to the  model Eq.~\eqref{eq:epsilon} through the ARPES data of Ref.~\onlinecite{Burganov2015} and DFT calculations.
Unstrained Sr$_2$RuO$_4$ and its close relative Ba$_2$RuO$_4$ have van Hove singularities of the $d_{xy}$  character(2D $xy$ band)  at the X and Y points slightly above (Sr$_2$RuO$_4$) or below (Ba$_2$RuO$_4$) the Fermi level (see the Supplementary Material (SM) section II.B). 
When applying uniaxial tensile strain in [100] direction on these quasi 2D ruthenates, one expects\cite{Hicks18042014,PhysRevB.93.054501} the bandwidth to decrease along [100] direction while behaving oppositely in the [010] direction. 
Our DFT calculations indeed predict the density of states of the $xz$ and $xy$ band to show similar amount of growth for small magnitude of uniaxial strain(see SM section II.A) although it is the $xy$ band that eventually reaches the van Hove singularity at X or Y at a large enough strain(see SM section II.A).
As for the biaxial strain, we predict 
Sr$_2$RuO$_4$ and Ba$_2$RuO$_4$ to reach the van Hove singularity at both X and Y points at the Fermi level under a tensile and compressive strain respectively (see Figure ~\ref{bands_strain} and SM section II.B), consistent with the experimental observations of Ref.~\onlinecite{Burganov2015} (see Fig.~\ref{FS}).
Moreover, we find this shift to the van Hove singularity to be driven by both the change in the bandwidth of the $xy$ band (see Fig.~\ref{bands_strain}) and the charge-transfer from the $xz$ and $yz$ bands (see SM section I). 
Nevertheless DFT consistently overestimates the Fermi velocities $v_F$ compared to ARPES, in particular that of the 2D band $xy$\cite{Burganov2015}. 

\begin{figure}[t]
\subfigure[]{
	\includegraphics[width=4cm]{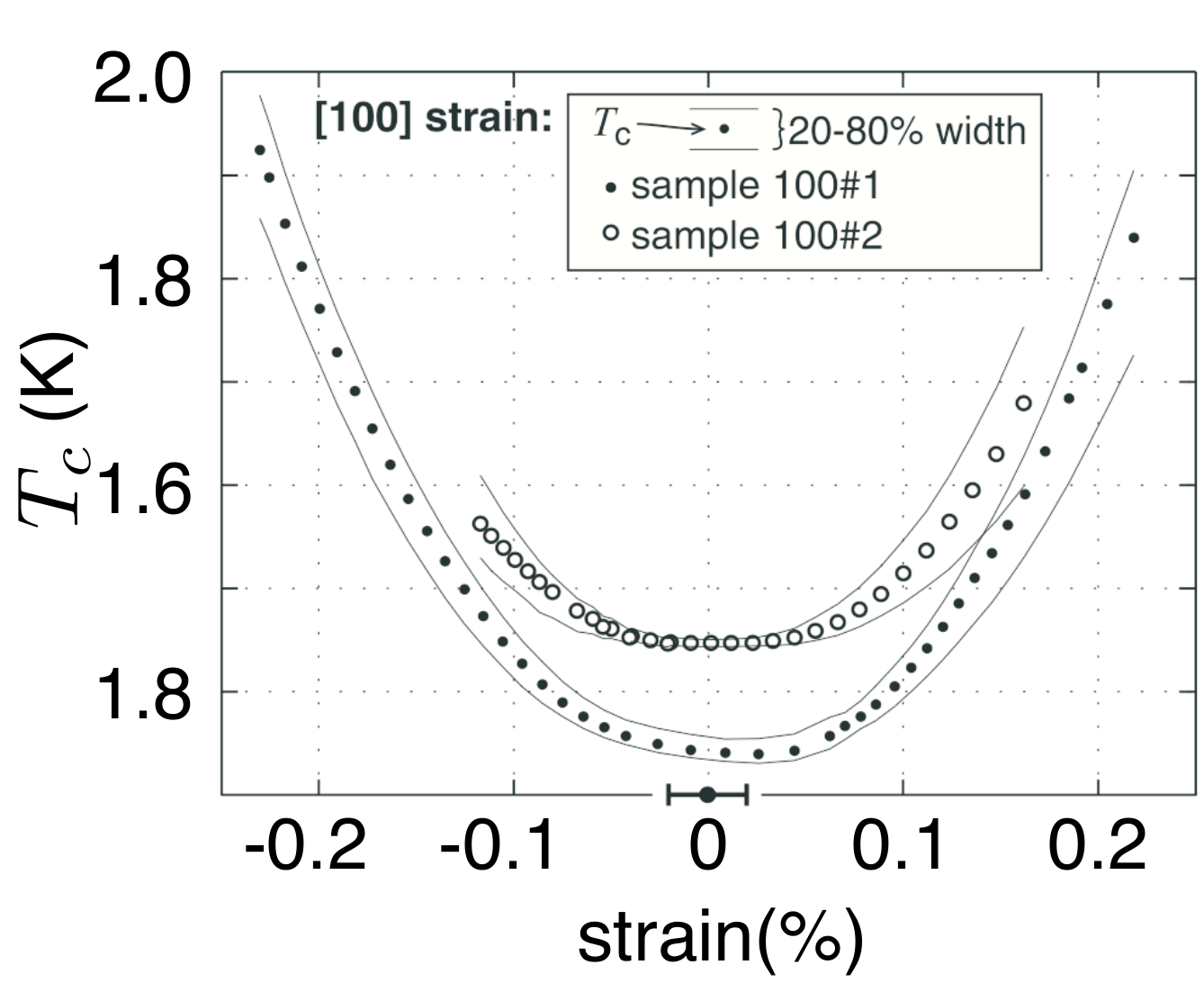}
	\label{fig:Tcstrain_exp}}
\subfigure[]{
	\includegraphics[width=4cm]{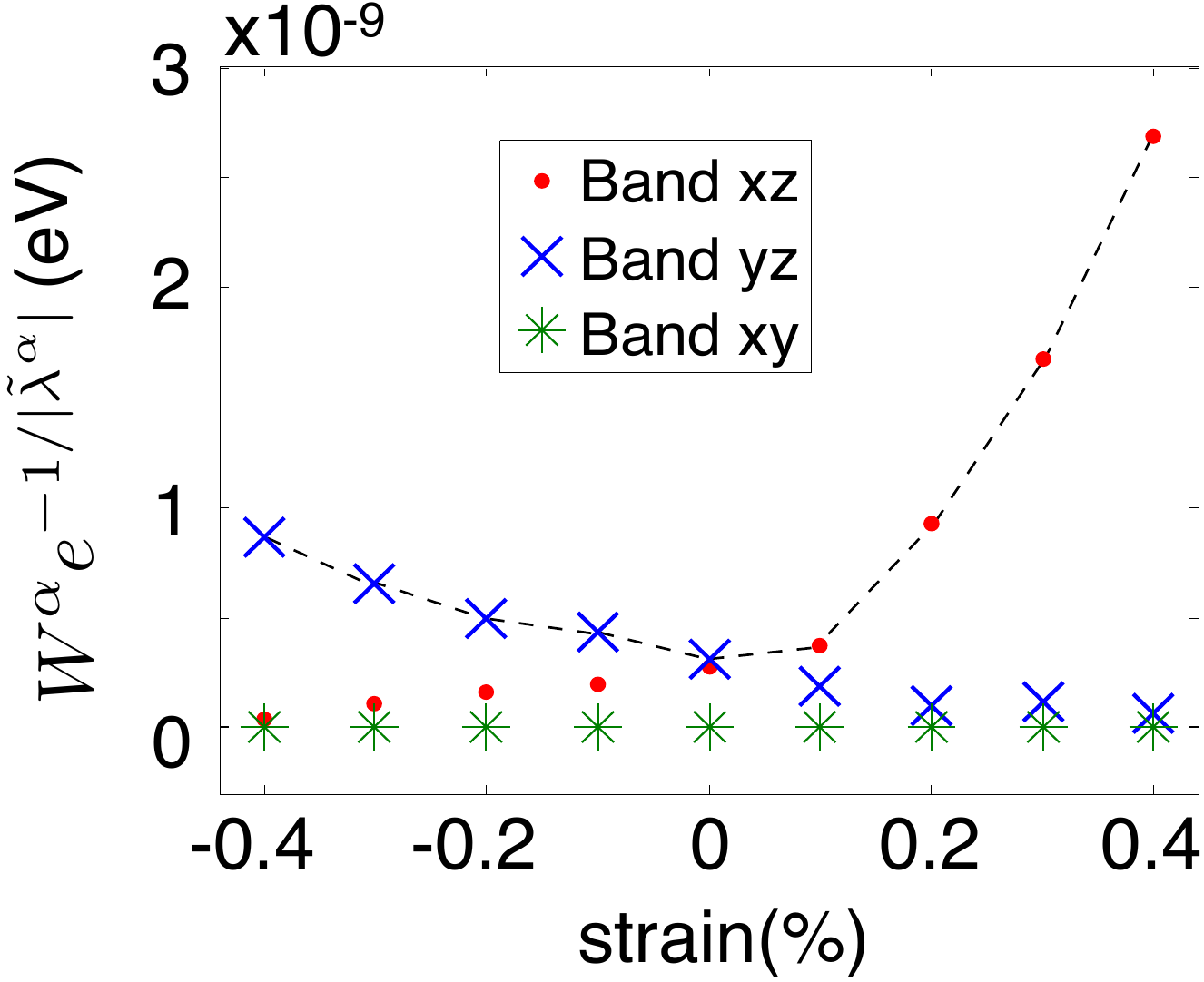}
	\label{fig:Tcstrain}}
\caption{(a)Measured $\rm{T_c}$ under both tensile($>$0) and compressive($<$0) uniaxial strain in [100] direction presented in Ref.~\onlinecite{Hicks18042014}. (b) Calculated quantity $W^{\alpha}e^{-1/|\tilde{\lambda}^{\alpha}|}$ under different amounts of uniaxial strain in the  [100] direction with $U=1$ eV. The black dashed line shows the expected transition temperature $\rm{T_c}$.}
\label{uniaxial}
\end{figure}

For completeness we now briefly review the two-step perturbative RG  approach\cite{PhysRevB.81.224505,PhysRevLett.105.136401} we adopt.
As a first step we numerically calculate the effective pairing vertices in different channels at some intermediate energy scale $E=\Lambda_0$ near the FS by integrating out higher energy modes down to $\Lambda_0$. 
To the one-loop order, the singlet and triplet effective pairing vertices $\Gamma_{s/t}^{\alpha}(\hat{k},\hat{k'})$ at energy $\Lambda_0$ are related to the repulsive bare interaction $U$ and the static particle-hole bubbles $\Pi_{ph}^{\alpha}(\vec{q})$ through
\begin{align}
\Gamma_s^{\alpha}(\hat{k},\hat{k'})=&U+U^2\Pi_{ph}^{\alpha}(\hat{q}=\hat{k}+\hat{k'}),
\label{eq:Gammas}
\end{align}
and 
\begin{align}
\Gamma_t^{\alpha}(\hat{k},\hat{k'})=&-U^2\Pi_{ph}^{\alpha}(\hat{q}=\hat{k}-\hat{k'}).\label{eq:Gammat}
\end{align}
Now the pairing tendency hosted by band $\alpha$ in each of the two pairing channels can be quantified by the most negative eigenvalue $\tilde{\lambda}_{s/t}^{\alpha}\equiv\lambda_{s/t}^{\alpha}(E=\Lambda_0)$ of a dimensionless matrix $g_{s/t}^{\alpha}(\hat{k},\hat{k'})$, which is a
 product of the density of states(DOS) $N^{\alpha}(\Lambda_0)\sim N^{\alpha}(0)$ and the normalized effective pairing vertices at the intermediate energy scale $\Lambda_0$:
\begin{align}
g_{s/t}^{\alpha}(\hat{k},\hat{k'})=N^{\alpha}(\Lambda_0)\sqrt{\frac{\bar{v_F}^{\alpha}}{v_F^{\alpha}(\hat{k})}}\Gamma_{s/t}^{\alpha}(\hat{k},\hat{k'})\sqrt{\frac{\bar{v_F}^{\alpha}}{v_F^{\alpha}(\hat{k'})}}.
\label{eq:gmatrix}
\end{align}
Here $\hat{k}^{(')}$ are the outgoing(incoming) momenta on the FS of band $\alpha$,  $v_F^{\alpha}(\hat{k})$ is the magnitude of Fermi velocity at $\hat{k}$, and $\frac{1}{\bar{v_F}^{\alpha}}\equiv\int\frac{d\hat{p}}{S_f^{\alpha}}\frac{1}{v_F^{\alpha}(\hat{p})}$ with $S_F^{\alpha}\equiv\int d\hat{p}$ being the FS `area' of orbital $\alpha$.
The second step is to study the RG flows of the most negative eigenvalues $\lambda_{s/t}^{\alpha}(E)$ for different channels ($\alpha$, $s/t$). Given the well known RG equations for the Cooper instability, 
 $\frac{d\lambda_{s/t}^{\alpha}}{dy}=-(\lambda_{s/t}^{\alpha})^2$ in terms of $y\equiv\log(\Lambda_0/E)$\cite{ShankarRG}, 
we can relate $\rm{T_c}$ to the most negative $\tilde{\lambda}_{s/t}^{\alpha}$ among all channels ($\tilde{\lambda}$) as
$\rm{T_c}\propto e^{-1/|\tilde{\lambda}|}$\cite{PhysRevLett.105.136401}.

\begin{figure*}[t]
\centering
\includegraphics[width=.6\textwidth]{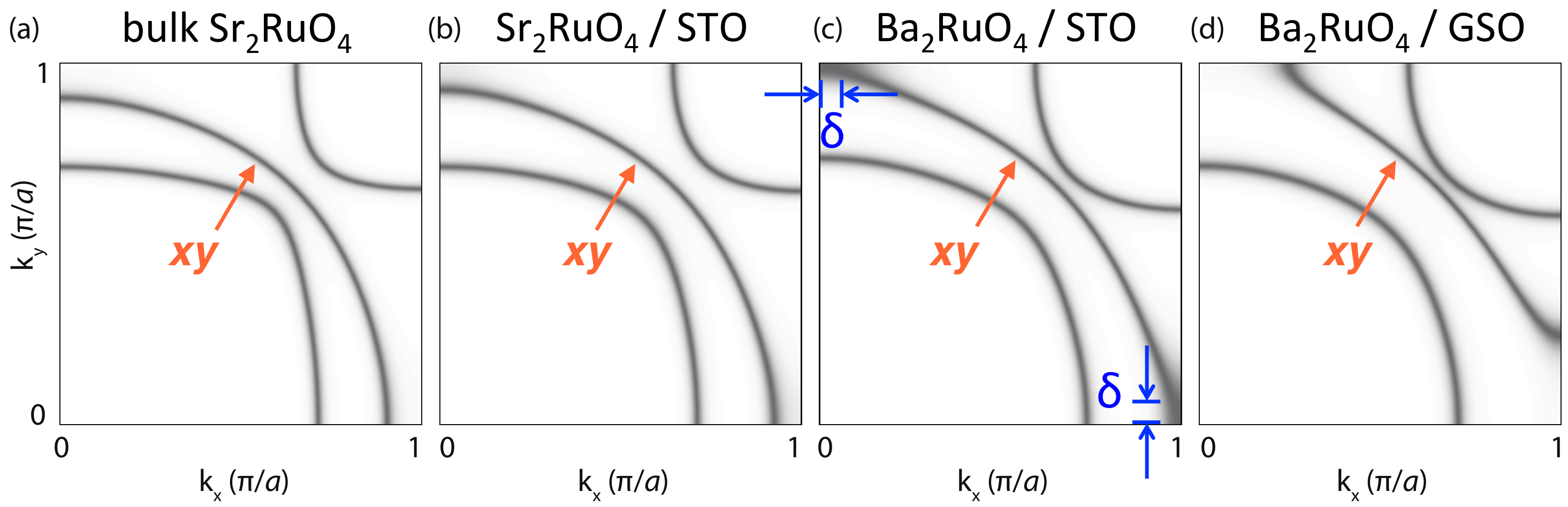}
\caption{The spectral functions extracted from the ARPES data in Ref.~\onlinecite{Burganov2015} for (a) the bulk $\rm{Sr_2RuO_4}$, (b) $\rm{Sr_2RuO_4}$/ STO, (c) $\rm{Ba_2RuO_4}$/ STO, and (d) $\rm{Ba_2RuO_4}$/ GSO respectively.  
The distance from the hole pockets to the van Hove singularities located at $\vec{k}$=($\pm\pi,0$) and ($0,\pm\pi$) are denoted by 
$\delta$ in (c). The parameterizations for the bulk $\rm{Sr_2RuO_4}$, $\rm{Sr_2RuO_4}$/ STO, $\rm{Ba_2RuO_4}$/ STO, and $\rm{Ba_2RuO_4}$/ GSO are ($t,t^{\perp},\mu_1,t',t'',\mu_2,t_{hyb}$) = (0.165, 0.0132, 0.178, 0.119, 0.0488, 0.176, 0.0215), (0.14, 0.0126, 0.148, 0.114, 0.0456, 0.171, 0.0224), (0.115, 0.0219, 0.112, 0.095, 0.0365, 0.1463, 0.0161), and (0.085, 0.0162, 0.074, 0.07, 0.0245, 0.11, 0.0136)
in the unit of eV respectively.}
\label{FS}
\end{figure*}

\textit{Uniaxially strained $Sr_2RuO_4$} $-$
\textcite{Hicks18042014} found the superconducting $\rm{T_c}$ of $\rm{Sr_2RuO_4}$
to enhance under both tensile and compressive uniaxial strain in [100] direction(see Fig.~\ref{uniaxial}(a)).  They then used phenomenological Ginzburg-Landau analysis to interpret that the enhancement of $\rm{T_c}$ was driven by the enhancement of density of states in one of the two quasi-one dimensional bands. Here, by determining the tight-binding parameters from DFT calculations under strain we gain insight into the interplay between strain and electronic structure. Further by feeding the the strained tight-binding parameters into the RG procedure, we can let our RG flow start from experimentally relevant short-distance physics. 

We then carried out the RG analysis to obtain the most negative eigenvalues $\tilde{\lambda}^{\alpha}_{s/t}$,
which are determined by the DOS $N^{\alpha}(0)$, band width $W^{\alpha}$, and the on-site repulsion $U$.
Fig.~\ref{uniaxial}(b) shows a quantity corresponding to $\rm{T_c}$ which involves the pairing tendency of band $\alpha$ quantified by the more negative eigenvalue between singlet and triplet channels $\tilde{\lambda}^{\alpha}\equiv \rm{min}(\tilde{\lambda}^{\alpha}_s,\tilde{\lambda}^{\alpha}_t)$ under different amounts of strain. 
We find that while the strain enhances the density of states of both 1D and 2D bands moderately (see SM section II.A), the strong pairing interaction of the 1D bands due to the antiferromagnetic fluctuation further amplifies the enhancement in the 1D pairing tendencies $|\tilde{\lambda}^{xz/yz}|$\footnote{We should note that DFT electronic structure significantly underestimate the density of states and the ferromagnetic fluctuation of the 2D band $xy$ in particular, which can affect the balance.}.
As the more dominant of the two 1D bands will onset the superconducting transition, our results imply the transition temperature $\rm{T_c}$ to follow  
the dashed curve in Fig.~\ref{uniaxial}(b)  as a function of tensile and compressive strain. 
Note that the estimated value of $U$ from first principle calculations is at the order of eV\cite{SROhubbardU}, which is beyond the weak coupling regime. Nonetheless, we set $U=1$ eV in  Fig.~\ref{uniaxial}(b) for illustrative purposes.  The so obtained  strain dependence of the $\rm{T_c}$ qualitatively captures the measured trend~\cite{Hicks18042014} shown in Fig.~\ref{uniaxial}(b). 

\textit{Biaxially strained ruthenates thin films} $-$
We now turn to the epitaxial ruthenate films under biaxial strain\cite{Burganov2015}. Biaxial strain has the advantage that it retains the tetragonal symmetry necessary for the onset of 
topologically non-trivial
$p_x+ip_y$ order parameter. Further, since X and Y points are approached simultaneously the van Hove singularity is expected to have a more substantial impact under biaxial strain(see SM section II. A and B). On the other hand, epitaxial strain can only access a discrete set of strain values and likely none will be precisely tuned to the van Hove point. But this may be a blessing since there are two theoretical issues when reaching the van Hove singularity. First, the van Hove points at X and Y points are forbidden from supporting a odd-parity triplet pairing by symmetry\cite{PhysRevB.92.035132,PhysRevB.89.144501}. Secondly, other ordering tendencies that can also benefit from the van Hove singularity can compete with superconductivity\cite{PhysRevB.86.020507,FuchunSROunistrain}. Hence by being close to a van Hove singularity without tuning into one, epitaxial biaxial strain may optimize triplet pairing.
 
Four representative samples we consider are the unstrained bulk $\rm{Sr_2RuO_4}$, a $\rm{Sr_2RuO_4}$ film grown on $\rm{SrTiO_3}$ ($\rm{Sr_2RuO_4}$/ STO),
a $\rm{Ba_2RuO_4}$ film grown on $\rm{SrTiO_3}$ ($\rm{Ba_2RuO_4}$/ STO), and
a $\rm{Ba_2RuO_4}$ film grown on $\rm{GdScO_3}$ ($\rm{Ba_2RuO_4}$/ GSO). 
See Fig.~\ref{FS}(a)-(d) for the associated spectral function of quasi-particles simulating the ARPES data, where the $xy$-band is electron-like in Fig.~\ref{FS}(a)-(b) and hole-like in Fig.~\ref{FS}(c)-(d) \footnote{The simulation included
 an additional hybridization term between the 1D bands with strength $V_{hyb}(\vec{k})=4t_{hyb}\sin k_x\sin k_y$  
extracted from the ARPES data\cite{Burganov2015}.}.
Interestingly, the DFT calculations consistently underestimate  the density of states and the Lindhard susceptibility at small  $\vec{q}$ of the $xy$-band: $N^{xy}(0)$ and $\Pi^{xy}_{ph}(\vec{q})$ at small $\vec{q}$. For our RG analysis we use the parameters 
$t_x=t_y\equiv t$, $t_x^{\perp}=t_y^{\perp}\equiv t^{\perp}$, and $t_x'=t_y'\equiv t'$ extracted from the ARPES data of Ref.~\onlinecite{Burganov2015}.

\begin{figure}[t]
	\includegraphics[width=.4\textwidth]{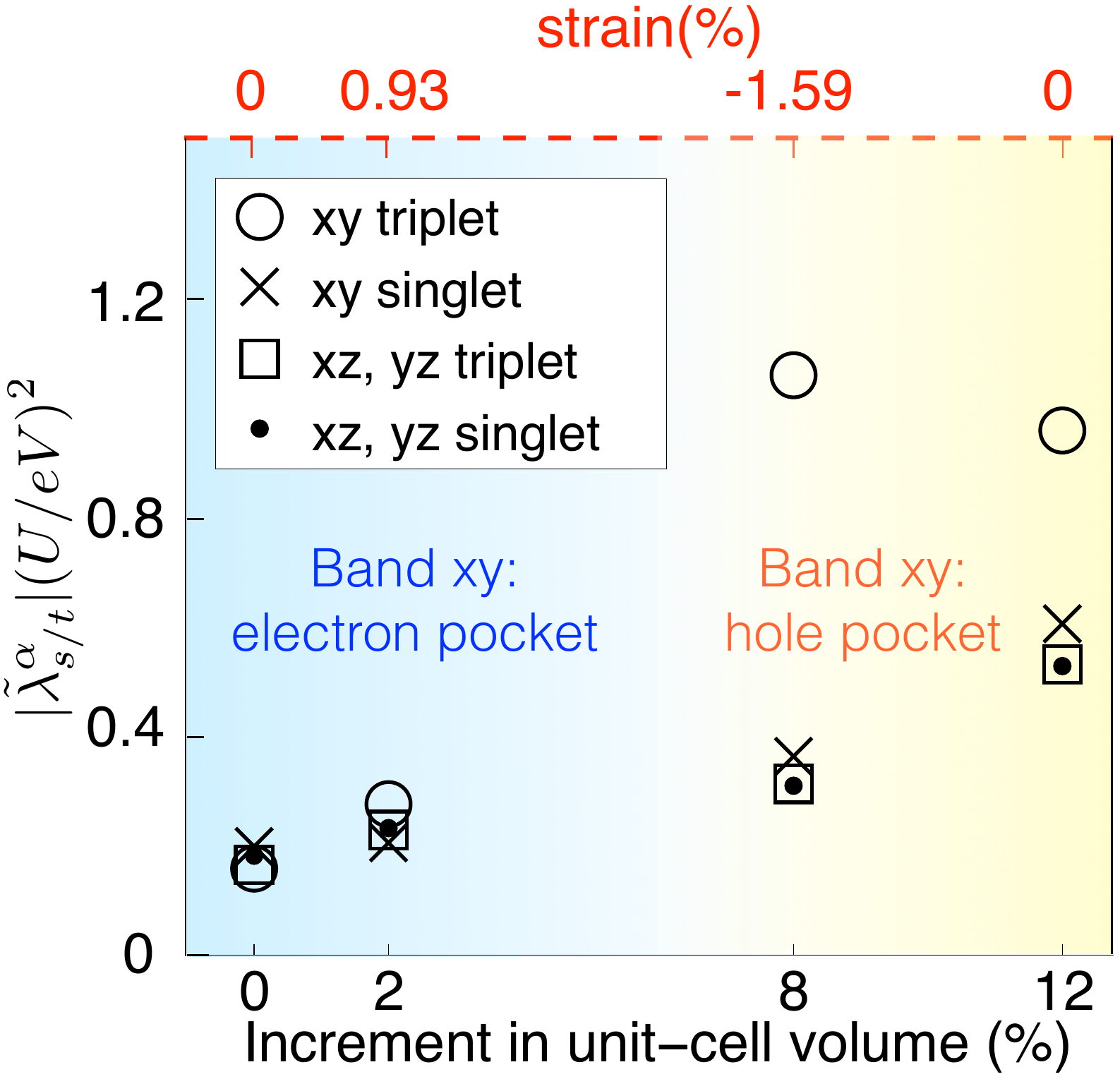}
\caption{The magnitudes of the most negative eigenvalues $\tilde{\lambda}_{s/t}^{\alpha}$ of different channels ($\alpha$, $s/t$) for the four representative samples. In the order of increasing volume of one unit cell, the ticks on the horizontal axis mark the four representative samples: the bulk $\rm{Sr_2RuO_4}$(0\%), and the films $\rm{Sr_2RuO_4}$/ STO(2\%), $\rm{Ba_2RuO_4}$/ STO(8\%), and $\rm{Ba_2RuO_4}$/ GSO(12\%). The percentage refers to the increase in the volume of one unit cell compared to that of the unstrained bulk $\rm{Sr_2RuO_4}$. The upper horizontal axis shows the in-plane strain of each $\rm{Sr_2RuO_4}$ and $\rm{Ba_2RuO_4}$ sample defined with respect to the bulk $\rm{Sr_2RuO_4}$ and $\rm{Ba_2RuO_4}$/ GSO respectively.}
\label{lambda}
\end{figure}

In Fig.~\ref{lambda} we show the resulting 
most negative eigenvalues $\tilde{\lambda}_{s/t}^{\alpha}$ of singlet and triplet channels hosted by each band $\alpha$ for the four representative samples.
Since the measured effect of strain on the band structures of the 1D bands is mild, the eigenvalues associated with the 1D bands do not change drastically. 
The tight competition between different channels of unstrained system\cite{PhysRevLett.105.136401,PhysRevB.89.220510} is lifted  as 
triplet pairing tendency of 2D band shoots up to become clearly leading instability in the vicinity of the Lifshitz transition. Moreover, this leading pairing tendency shows a striking non-monotonic
dependence on the strain with significantly improved pairing tendency in film $\rm{Ba_2RuO_4}$/ STO. Importantly, as the $\rm{Ba_2RuO_4}$/ STO film is slightly away from the actual van Hove singularity by a small distance $\delta$(see Fig.~\ref{FS}(c)) triplet pairing is allowed by symmetry. 

The significant enhancement in the triplet pairing tendency of the 2D band in the $\rm{Ba_2RuO_4}$/ STO film is due to the conspiracy between the enhanced DOS of the 2D band and associated enhancement in the ferromagnetic fluctuation in the measured band structure
that enters the triplet pairing vertex through the bare particle-hole bubble $\Pi^{xy}_{ph}(\vec{q}=0)$. Interestingly, although the singlet pairing tendency of the 2D band also benefits from the enhanced DOS of the 2D FS near the van Hove singularity, the antiferromangetic fluctuation which facilitates the singlet pairing does not benefit from the proximate van Hove singularities as much due to the lack of perfect nesting. 

\textit{Summary} $-$ 
In summary, we theoretically investigated how strain-driven changes of band structure should impact the superconducting instabilities in ruthenates. 
Considering the effect of mild uniaxial strain of the degree achieved in Ref.~\onlinecite{Hicks18042014}, we confirmed our approach of using the strained bandstructure as an input to the RG calculation qualitatively reproduce the observed $\rm{T_c}$ -dependence on the lattice strain.
We then noted by order few $\%$strain the FSs can be altered sufficiently to come close to the nearby van Hove singularity. As such  degree of biaxial strain has been achieved by some of us\cite{Burganov2015} and shown to result in van Hove singularity in the 2D band, we used the band structure 
extracted from ARPES data as the input to the weak-coupling RG procedure and predicted 
triplet superconductivity with enhanced $\rm{T_c}$
to be driven predominantly by 2D bands near van Hove singularity. 
In order to test our predictions, the film purity\cite{SROscDisorder} and structural order\cite{SROscDefect,SROscCrystalDefect} 
need to improve. Recent success in growing superconducting $\rm{Sr_2RuO_4}$ thin films\cite{SROscfilm} makes us optimistic that point defects and extended defects of strained films can be sufficiently reduced. 

It is important to note that in the proposed strategy of engineering Fermi surface and using the resulting band structure as an input to an RG procedure, the aspects of results that are of great interest such as pairing channel and the $T_c$ are non-universal aspects that are sensitive to microscopic details. As we propose to use this very sensitivity to engineer a desired superconductor with the advancement of experimental capabilities\cite{Hicks18042014,Burganov2015,HicksVHS}, we should also stress the importance of basing the microscopic model to the measurement of the actual band structure. 

{\it Acknowledgement --}
The authors are grateful to Andrew Mulder, Srinivas Raghu, Andrey Chubukov, Andy MacKenzie, Thomas Scaffidi, Fuchun Zhang, Steve Kivelson, Ronny Thomale, and Hong Yao for helpful discussions. After completion of this work, we have become aware of two preprints Ref.~\onlinecite{HicksVHS} and Ref.~\onlinecite{FuchunSROunistrain} focusing on uniaxial strain effects on Sr$_2$RuO$_4$. \textcite{HicksVHS} experimentally achieved the necessary uniaxial strain to access the van Hove singularity and compared the results with theoretical approach similar to that used in this paper but with somewhat different microscopic model. \textcite{FuchunSROunistrain} used a functional renormalization group approach on a single band model focusing on the 2D band to study the competition between superconductivity and spin density waves under uniaxial strain approaching van Hove. 
This work was supported by the Cornell Center for Materials Research with funding from the NSF MRSEC program (DMR-1120296).  Y-TH and E-AK acknowledge support from NSF grant no. DMR-0955822. WC was supported by NSF grant no. DMR-1265593.

%

\maketitle
\begin{center}
{\bf\normalsize{SUPPLEMENTARY MATERIAL}}
\end{center}
\section{Effects of strain on bandwidth and occupation}
When either uniaxial or biaxial tensile strain is applied to the ruthenates, the bond length along the direction of strain increases 
which causes band-flattenings in the same direction (see Fig. 1 of the main text).
On the other hand, a dramatic change in the shape of the bands is at the same time accompanied by a charge-transfer among different bands. 
In Fig.~\ref{crystal_field} we show the $d$-orbital crystal field splitting corresponding to the RuO$_6$ octahedra in $\rm{Sr_2RuO_4}$ before and after applying strain (for comparison we also include the splitting corresponding to a perfect octahedral environment). 
In the biaxial case, tensile strain enlarges both in-plane lattice parameters while simultaneously reducing the out-of-plane one. The resulting octahedron is then less elongated and the energy splitting between orbital $xy$ and orbital $xz$, $yz$ is reduced. Consequently the occupation of the $xy$ level increases.
Such charge-transfer together with the flattening of the $xy$ band along the strain direction could bring the quasi-2D Fermi surface which is already close to van Hove singularity in the absence of strain to one or both of the van Hove points $X=(\pi,0)$ and $Y=(0,\pi)$.  
Similar effects occur in $\rm{Ba_2RuO_4}$ under an opposite direction of strain due to the fact that the Fermi level lies above the van Hove points instead of below (see Fig. 1 of main text).

\section{DFT calculation on strained ruthenates}
In order to gain a deeper insight into the interplay between strain and electronic structure in ruthenates we perform DFT calculations by systematically varying the applied amount of uniaxial and biaxial strain. We use the PBEsol exchange-correlation functional as implemented in VASP\cite{PhysRevB.54.11169,PhysRevB.50.17953} with a plane wave basis cutoff of 520 eV. For structural and static calculations we use a 8x8x4 and 12x12x12 sampling of the Brillouin zone, respectively.
Full structural relaxations are performed on $\rm{Ba_2RuO_4}$ and $\rm{Sr_2RuO_4}$ and the optimized cells thus obtained are the unstrained unit cells we later use in uniaxial and biaxial calculations. The resulting lattice constants are in both cases in good agreement with experiment:  the obtained values are $a=3.947 \rm{\AA}$ (exp. $3.990 \rm{\AA}$), $c=13.417 \rm{\AA}$ (exp. $13.430 \rm{\AA}$) for $\rm{Ba_2RuO_4}$, and $a=3.831 \rm{\AA}$ (exp. $3.871 \rm{\AA}$) , $c=12.731 \rm{\AA}$ (exp. $12.739 \rm{\AA}$) for $\rm{Sr_2RuO_4}$\cite{Burganov2015}.
Despite the negligible underestimation of the experimental value for the $c$ lattice constant, we find the calculated in-plane lattice constant $a$ to be underestimated by a 1\% for both ruthenates, which leads to an artificial elongation in the RuO$_6$ octahedron.
Moreover, while the Fermi velocities of all three bands $xz$, $yz$, $xy$ are underestimated compared to the ARPES data,
in particular we find the Fermi velocity of band $xy$ to be underestimated the most which agrees with the observation made in Ref.~\onlinecite{Burganov2015}.
These underestimation in turns reduces the $xy$ band occupations and the DOS at the Fermi level for this band.


\subsection{Uniaxial strain}
\begin{figure}[]
\vspace{1cm}
\includegraphics[width=.5\textwidth]{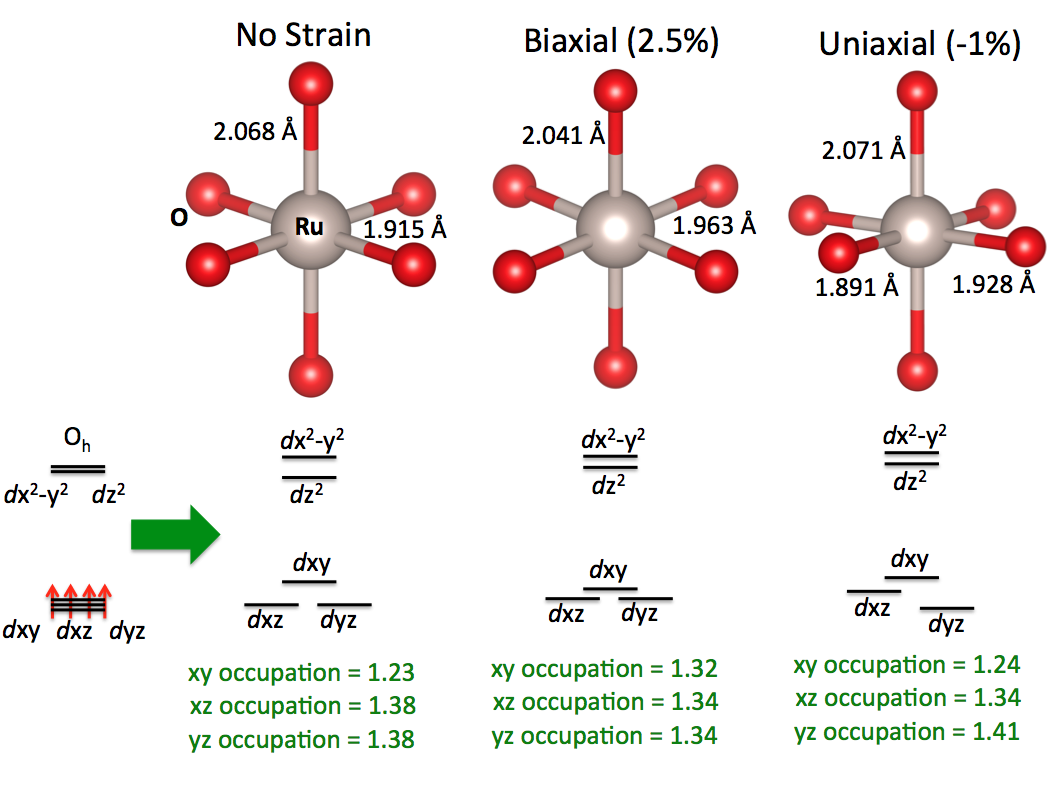}
\caption{Crystal field splitting of energy levels in $\rm{Sr_2RuO_4}$: unstrained, under tensile biaxial strain and under compressive [100] uniaxial strain. Octahedral symmetry is broken in the three cases. In the biaxial case, changes in the bond lengths decrease the energy difference between $xy$ and $xz,yz$ levels, increasing the $xy$ occupation. For uniaxial case, bond lengths are reduced along the [100] direction and increased along [010], as a result the $xz$ and $yz$ levels are shifted up and down, respectively.}
\label{crystal_field}
\end{figure}

\begin{figure}[]
\vspace{1cm}
\includegraphics[width=.3\textwidth]{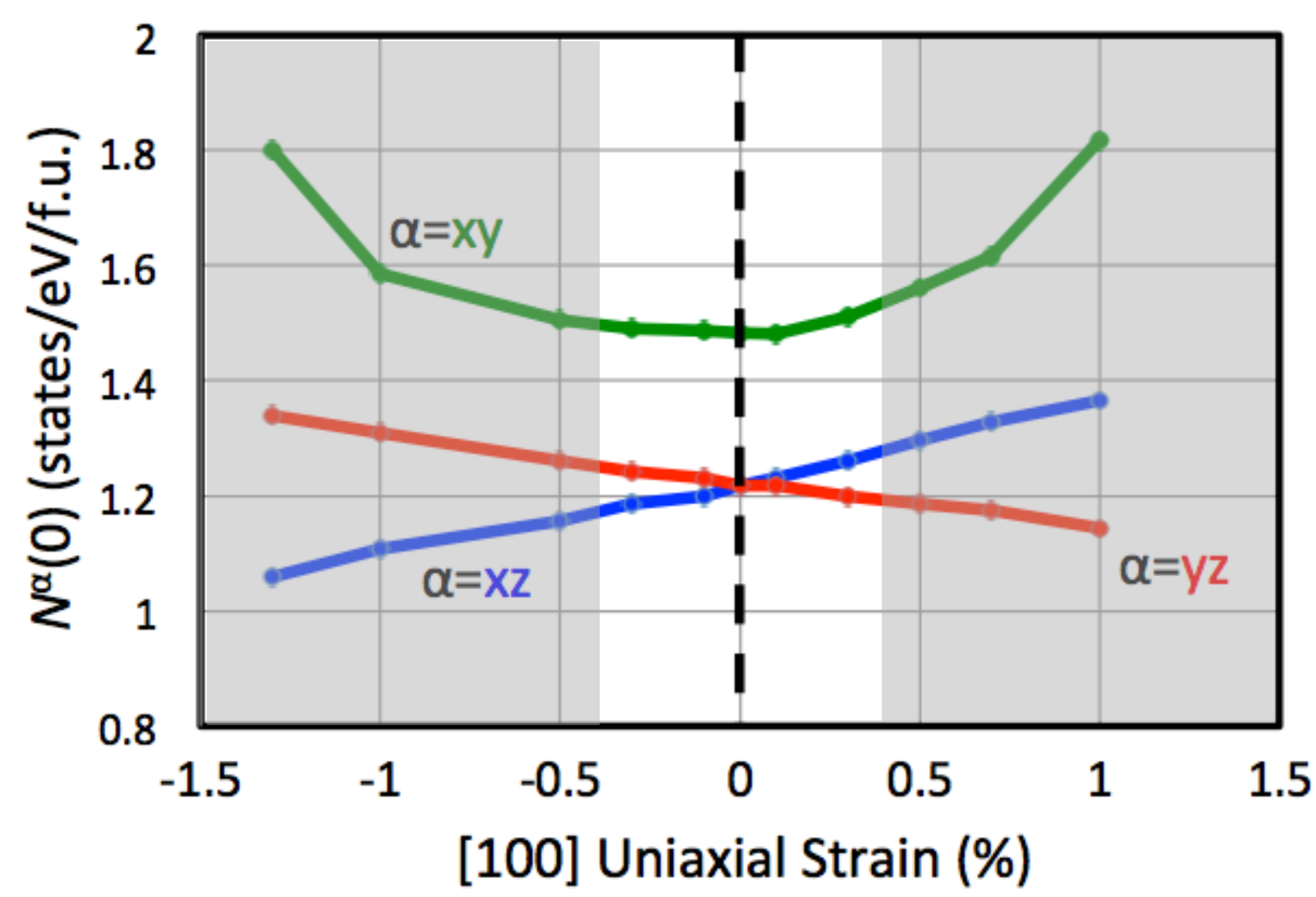}
\caption{The density of states at the Fermi level for $xy$, $xz$ and $yz$ bands under tensile($>$0) and compressive($<$0) uniaxial strain. The unshaded regime is the regime with small strain magnitude which corresponds to Fig. 2(b) in main text.}
\label{dos_fermi_uniaxial}
\end{figure}

\begin{figure}[]
\vspace{1cm}
\includegraphics[width=.5\textwidth]{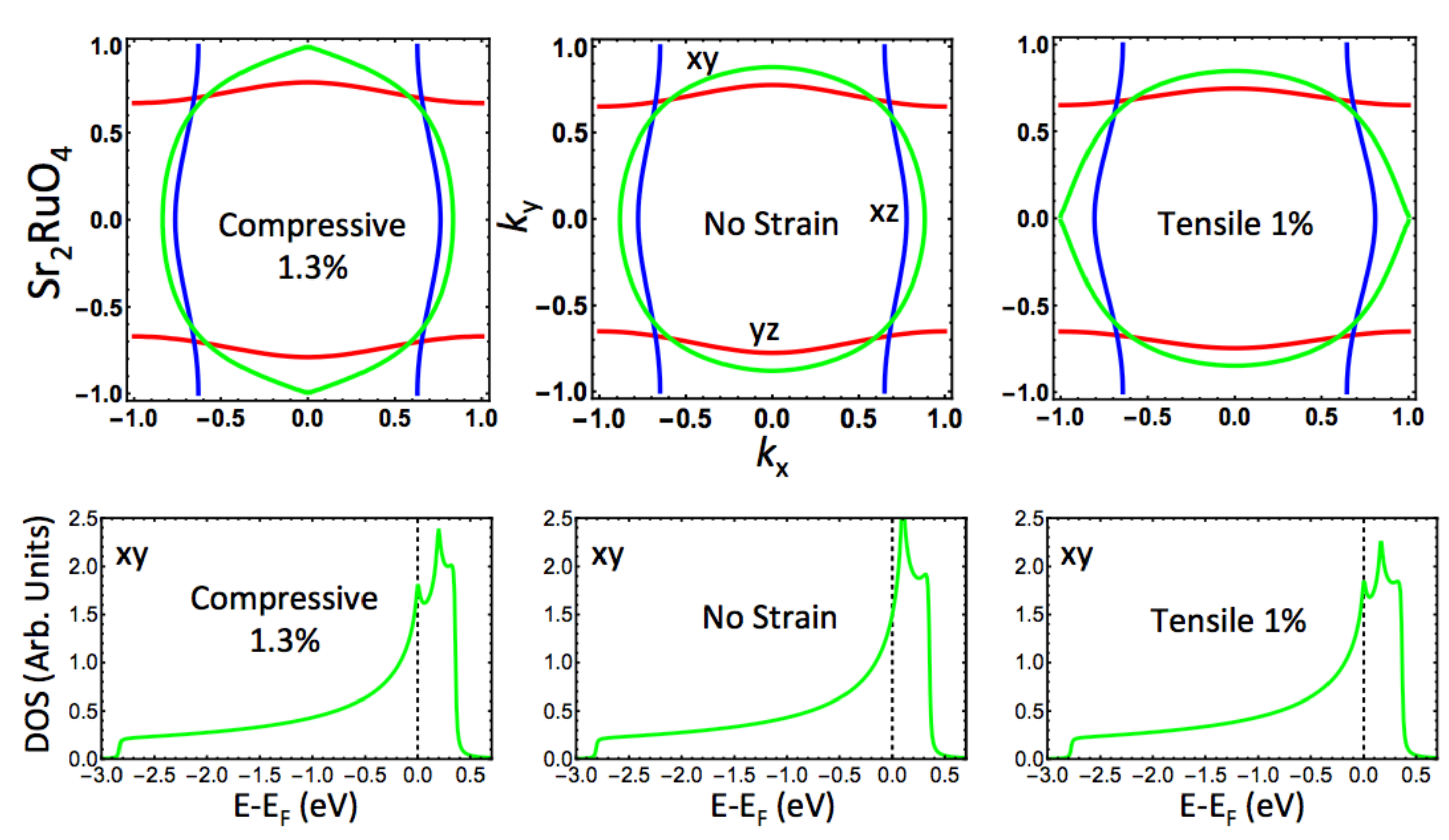}
\caption{FSs  at $k_z=0$ for $\rm{Sr_2RuO_4}$ (top panel) and the corresponding DOS for the $xy$ band (bottom panel) as a function of [100] uniaxial strain. The uniaxial strain lowers the symmetry from $D_{4h}$ to $D_{2h}$. As a result, while the peak in the DOS of $xy$ band due to the van Hove point $X=(\pi,0)$ sits at the Fermi level, the peak due to the van Hove point $Y=(0,\pi)$ lies above the Fermi level by around 200 meV. Units of $k_x$ and $k_y$ are $\pi/a$ and $\pi/b$ respectively, where $a$ and $b$ are the in-plane lattice constants.}
\label{uniaxial_fs_dos}
\end{figure}
We relax the internal structural degrees of freedom and transverse lattice constants by keeping the [100] direction fixed at the desired strain amount. The Fermi surfaces (FSs) and dispersions obtained from DFT are then used to fit the parameters for the tight-binding model in Eq.~(2) in the main text. 

In Fig.~\ref{dos_fermi_uniaxial} we show the evolution of the band-projected density of states (DOS) at Fermi level ( $N^\alpha(0)$) as a function of  [100] uniaxial strain in $\rm{Sr_2RuO_4}$. As bond lengths along the $x$-direction are reduced, bond lengths along $y$ increase. Consequently $yz$ bands become less dispersive and the number of states at Fermi level is increased while it is reduced for $xz$. At the same time occupation for the $xz$ bands decreases, while increasing for $yz$ (see Fig.~\ref{crystal_field}). Note that though the DOS of the $xy$ band starts to thrive at a large amount of strain due to the van Hove singularity, the growths in DOS of $xz$($yz$) and $xy$ bands under a small amount of tensile(compressive) strain are of similar magnitudes as shown in the unshaded area. 
This small strain regime is the regime where we investigate the superconducting tendencies in Fig. 2(b) of the main text. 

In our calculation, the $xy$ band meets the van Hove points $X$ and $Y$ at the Fermi level for a [100] compressive strain of 1.3\% and for a [100] tensile strain of  1\% (see the top panel of Fig.~\ref{uniaxial_fs_dos}). The larger value for [100] compressive strain is due to the in-plane Poisson ratio ($\approx$ 0.4)\cite{PoissonRatioSRO} of $\rm{Sr_2RuO_4}$. 
From the $xy$ band DOS shown in the bottom panel of Fig.~\ref{uniaxial_fs_dos}, we notice the peak due to van Hove singularities in the unstrained case splits into two peaks under strain. This is a consequence of uniaxial strain reducing the symmetry from $D_{4h}$ to $D_{2h}$. 

\begin{figure}[]
\vspace{1cm}
\includegraphics[width=.5\textwidth]{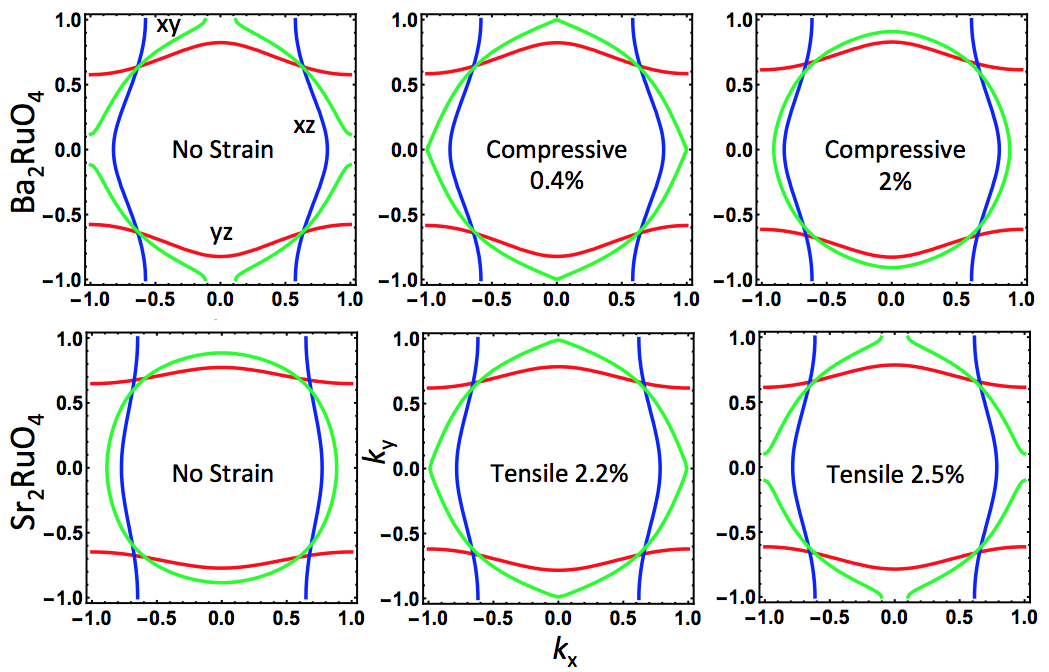}
\caption{FSs at $k_z=0$ for $\rm{Ba_2RuO_4}$ (top panel) and $\rm{Sr_2RuO_4}$ (bottom panel) as a function of strain obtained by fitting the tight-binding model to DFT data. While the Fermi level of $\rm{Sr_2RuO_4}$ approaches the van Hove points $X=(\pi,0)$ and $Y=(0,\pi)$ under a tensile strain, the opposite trend is predicted for $\rm{Ba_2RuO_4}$. $k_x$ and $k_y$ are in the unit of $\pi/a$ where $a$ is the in-plane lattice constant.}
\label{FS_dft_strain}
\end{figure}

\begin{figure}[]
\includegraphics[width=.5\textwidth]{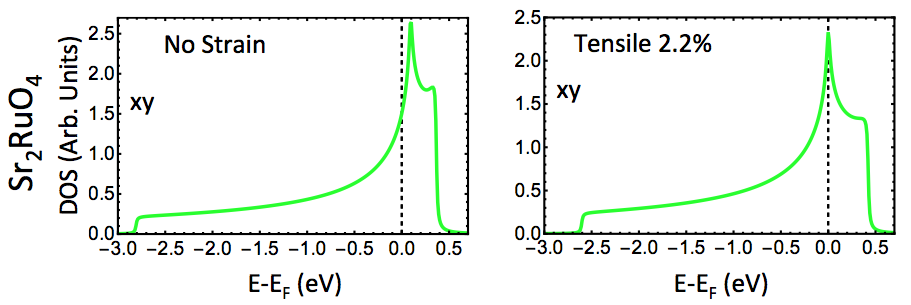}
\caption{The density of states for $\rm{Sr_2RuO_4}$ before and after applying the biaxial strain.}
\label{sro_xy_dos}
\end{figure}

\subsection{Biaxial strain}
We perform DFT calculations for a wide range of values of tensile and compressive strain by fixing the in-plane lattice constants and letting all internal and out-of-plane lattice constant to relax. We then fit the tight-binding model presented in the main text to our DFT data. 
As illustrated in Fig. 1 in the main text, in the case of $\rm{Sr_2RuO_4}$ the hopping parameters decrease and the bands become flatter with tensile strain, thus approaching the van Hove singularity. Due to the simultaneous increase of in-plane bond lengths together with a decrease of the out-of-plane ones, $xy$ energy levels move downwards while $xz$ and $yz$ are shifted up (Fig.~\ref{crystal_field}). As mentioned in section I, this results in a enhancement of the $xy$ occupation which further contributes to reaching the van Hove singularity.
Our calculations predict a peak in the DOS of the $xy$ band  due to the Van Hove singularities at both $X$ and $Y$ at the Fermi level for a tensile strain of $\approx$ 2.2 \% (lower panel in Fig.~\ref{FS_dft_strain} and Fig.~\ref{sro_xy_dos}). 
Contrarily, for $\rm{Ba_2RuO_4}$ it requires a compressive strain of $\approx$ 0.4 \% (upper panel in Fig.~\ref{FS_dft_strain}). 
This is due to the fact that in $\rm{Ba_2RuO_4}$ the $xy$ band lies below the Fermi level and thus a compressive strain needs to be applied in order to increase the hopping parameters and broaden the band (see Fig.~1 in the main text). 
One important difference from the uniaxial strain case is that $D_{4h}$ symmetry is preserved under biaxial strain such that the $xy$ band FS can meet both van Hove points simultaneously. 
Thus the DOS of $xy$ band only has a single peak due to the van Hove singularity with a much higher intensity (see Fig.~\ref{sro_xy_dos}) than the small split peak at the Fermi level of uniaxially strained $\rm{Sr_2RuO_4}$ (see the bottom panel of Fig.~\ref{uniaxial_fs_dos}). 

\end{document}